\documentclass{article}

\usepackage{amssymb}
\usepackage{algorithmic}
\usepackage{epsfig}

\newtheorem{theorem}{Theorem}[section]

\newtheorem{algorithm}[theorem]{Algorithm}

\newcommand{\CC}{\mathbb C}
\newcommand{\bfx}{{\bf x}}
\newcommand{\f}{{\bf f}}
\newcommand{\LL}{{\bf L}}
\newcommand{\HH}{{\bf H}}
\newcommand{\zero}{{\bf 0}}

\begin{document}

\title{Decomposing Solution Sets of Polynomial Systems: \\
       A New Parallel Monodromy Breakup Algorithm\thanks{This material
    is based upon work
    supported by the National Science Foundation
    under Grant No.\ 0134611 and Grant No.\ 0410036.}}

\author{Anton Leykin\thanks{
leykin@math.uic.edu.
http://www.math.uic.edu/{\~{}}leykin.} and
Jan Verschelde\thanks{
jan@math.uic.edu. http://www.math.uic.edu/{\~{}}jan.}
}

\date{\today}

\maketitle

\begin{abstract}
We consider the numerical irreducible decomposition of a positive
dimensional solution set of a polynomial system into irreducible factors.
Path tracking techniques computing loops around singularities connect
points on the same irreducible components.  The computation of a linear
trace for each factor certifies the decomposition.
This factorization method exhibits a good practical performance on
solution sets of relative high degrees.

Using the same concepts of monodromy and linear trace, we present
a new monodromy breakup algorithm.  It shows a better performance
than the old method which requires construction of permutations of
witness points in order to break up the solution set.
In contrast, the new algorithm assumes a finer approach allowing
us to avoid tracking unnecessary homotopy paths.

As we designed the serial algorithm keeping in mind distributed
computing, an additional advantage is that its parallel version
can be easily built.  Synchronization issues resulted in a
performance loss of the straightforward parallel version of the
old algorithm.  Our parallel implementation of the new approach
bypasses these issues, therefore, exhibiting a better performance,
especially on solution sets of larger degree.



\end{abstract}

\section{INTRODUCTION}

\subsection{Problem Statement}

As polynomial equations emerge more and more often in various
fields of science and engineering, the question of simplification
of polynomials and polynomial systems becomes of the most
importance. How can we simplify? One way to understand better the
solution set of a polynomial is to factor it; equivalently, in
case of a polynomial system we talk about finding an
\emph{irreducible decomposition} of its solution set,
a central problem in {\em numerical algebraic geometry}~\cite{SW96,SW05}.

A widely known family of polynomial systems used for benchmarking
is ``cyclic $n$-roots'', which arose in Fourier analysis~\cite{Bjo85,Bjo89}.
The case $n=4$ is our running example:

\begin{equation}  \label{eq_cyclic4}
\left\{
  \begin{array}{rcl}
     x_1 + x_2 + x_3 + x_4 & = & 0 \\
     x_1 x_2 + x_2 x_3 + x_3 x_4 + x_4 x_1 & = & 0 \\
     x_1 x_2 x_3 + x_2 x_3 x_4 + x_3 x_4 x_1 + x_4 x_1 x_2 & = & 0 \\
     x_1 x_2 x_3 x_4 - 1 & = & 0.
  \end{array}
\right.
\end{equation}
This system has a one dimensional solution component of degree four,
which becomes obvious by the substitution $x_3 = -x_1$ and
$x_4 = -x_2$.  After this substitution, the first three equation
vanish and the last equation simplifies to $x_1^2 x_2^2 - 1$.
Since $x_1^2 x_2^2 - 1 = (x_1 x_2 - 1)(x_1 x_2 + 1)$, the curve
of degree four factors in two irreducible quadrics.

We denote systems of polynomial equations by
$\f(\bfx)=(f_1(\bfx),\ldots,f_m(\bfx))=\zero$,
where $f_i\in \CC[\bfx]=\CC[x_1,\ldots,x_n]$ for all $i$.
Very often, the coefficients are known with limited accuracy.
The solution set $V$ to $\f(\bfx) = \zero$ is naturally organized
into pure dimensional solution sets $V = [V_0,V_1,\ldots,V_n]$,
where ${\rm dim}(V_k) = k$.
A numerical representation of a pure dimensional solution set $V_k$ is
a \emph{witness set}~\cite{SVW1}~\cite{SW05}, which consists of
\begin{enumerate}
\item the polynomials $f_i$ $(i=1,\ldots,m)$;

\item $k$ linear equations $\LL(\bfx)=(L_1(\bfx),\ldots,L_k(\bfx))=\zero$
      with generic coefficients describing $k$ generic hyperplane
      slices;

\item a list $W$ of $\deg(V_k)$ solutions to the system
      $\f(\bfx)=\LL(\bfx)=\zero$.

\end{enumerate}
By the generic choice of the coefficients of the $\LL(\bfx) = \zero$,
the $k$ hyperplanes defined by $\LL$ cut out exactly as many isolated
regular solutions on $V_k$ as~$\deg(V_k)$.

Notice how the treatment of positive dimensional solution sets is
reduced to dealing with collections of generic points.
Using slack variables we reduce overdetermined polynomial systems
to systems with as many variables as unknowns~\cite{SV00}.
For a system like cyclic 4-roots in~(\ref{eq_cyclic4}), we add
one slack variable $z$ to the system:
\begin{equation} \label{eq_cyclic4slack}
   \left\{
      \begin{array}{r}
     x_1 + x_2 + x_3 + x_4 + b_1 z = 0~ \\
     x_1 x_2 + x_2 x_3 + x_3 x_4 + x_4 x_1 + b_2 z = 0~ \\
     x_1 x_2 x_3 + x_2 x_3 x_4 + x_3 x_4 x_1 + x_4 x_1 x_2 + b_3 z = 0~ \\
     x_1 x_2 x_3 x_4 - 1 + b_4 z = 0~ \\
     c_0 + c_1 x_1 + c_2 x_2 + c_3 x_3 + c_4 x_4 + z = 0,
      \end{array}
   \right.
\end{equation}
where the coefficients $b_1$, $\ldots$, $b_4$, $c_0$, $c_1$, $\ldots$,
$c_4$ are randomly chosen complex numbers.  The extra linear equation
reduces the dimension of the solution set by one and we find generic
points on the curve as regular solutions with~$z=0$.
For the cyclic 4-roots system the witness set contains four generic
points, as the degree of the curve equals four.

The main question now is:  given a positive dimensional
solution set $V$, can we find its decomposition
into the irreducible components?
In the language of witness sets
this interprets as: given a witness set $(\f,\LL,W)$ of $V$, can
we find a decomposition $W = W^{(1)} \sqcup \ldots \sqcup W^{(r)}$
such that for all $i$ the witness set $(\f,\LL,W_k^{(i)})$
represents an irreducible component of $V$?

Finding polynomial time algorithms for the factorization
of multivariate polynomial with approximate coefficients was posed
in~\cite{Kal00} as one of the challenges in symbolic computation.
This challenge received a lot of
attention~\cite{Che04,CG05,CGHKW01,CGKW02,GR01,GR02,GKMYZ04,HWSZ00,Sas01,SVW8};
see~\cite{CG05} for a nice description of recent methods.

\subsection{Numerical Homotopies define Loops around Singularities}

In~\cite{SVW3}, a new numerical algorithm using homotopy continuation
methods was proposed to decompose a positive dimensional solution set
into irreducible factors.  Linear traces were proposed in~\cite{SVW4}
to certify a numerical irreducible decomposition.
The implementation~\cite{SVW7} was adjusted to the important special
case of factoring one single multivariate complex polynomial in~\cite{SVW8}.

In this section we outline the idea of exploiting {\em monodromy}
using homotopies to define loops around singularities.
Assume two witness sets $(\f,\LL_1,W_1)$ and $(\f,\LL_2,W_2)$
represent the same positive dimensional irreducible component~$V$.
Consider the system $\HH_{\gamma, \LL_1, \LL_2}(\bfx,t)$:
\begin{equation}
\left\{
  \begin{array}{rcl}
    \f(\bfx) &=& \zero; \\
    (1-t)\LL_1(\bfx) + \gamma t \LL_2(\bfx)&=& \zero.
  \end{array}
\right. \ \ \ \ \ (t\in[0,1])
\end{equation}
where $\gamma$ is a generic nonzero complex number.
Then, due to the generic choice of $\gamma$, for a fixed value of $t$
the solutions to $\HH_{\gamma, \LL_1, \LL_2}(\bfx,t)$ are all
isolated.  In particular, these are $W_1$ for $t=0$ and $W_2$ for $t=1$.

Tracking solutions of $\HH_{\gamma, \LL_1, \LL_2}(\bfx,t)$ as $t$
varies from $0$ to $1$ defines a 1-to-1 map
$\phi_{\gamma, \LL_1, \LL_2}: W_1 \to W_2$.
If the composition of two such maps defines a loop around a
singularity for some~$t$, then a permutation of the points of
a witness set is obtained, in particular:
\begin{equation}
  \pi_{\gamma, \LL_1, \LL_2}
  = \phi_{\gamma_2, \LL_2, \LL_1} \circ \phi_{\gamma_1, \LL_1, \LL_2}
\end{equation}
is a permutation of $W_1$.  All permutations that arise in this
fashion form a subgroup of the symmetry group acting on $W_1$ and
the orbits of this action are witness sets that represent
irreducible components.

The idea to exploit monodromy first appeared in a theoretical
complexity study~\cite{BCGW93}.  Although our approach does not
need to know the precise location of the singularities, one could
as in~\cite{DvH01} compute those for algebraic curves, see also
the command {\tt algcurves[monodromy]} in Maple.

The algorithm in~\cite{SVW3} collects points connected by loops
into the same witness sets which converge to numerical representations
of the irreducible components.  In~\cite{SVW4}, a stop criterion for
this algorithm was presented, using the linear trace.  We explain this
trace test on a system like cyclic 4-roots.  Note our program only works
with generic points obtained as solutions of~(\ref{eq_cyclic4slack}) and
does not have a symbolic polynomial of degree four.
Via a generic projection we map the points in 4-space down to the plane.
If two of the four points to belong to the same irreducible component,
there must exist a quadratic polynomial $p(x,y)$ vanishing at those two
points and at any point of the quadratic irreducible factor.
The linear trace is then defined rewriting $p(x,y)$ as $p(x,y(x))$:
\begin{eqnarray}
  p(x,y(x)) \!\! & = & \!\! (y - y_1(x))(y - y_2(x)) \\
            \!\! & = & \!\! y^2 - (y_1(x) + y_2(x)) y + y_1(x) y_2(x) \\
            \!\! & = & \!\! y^2 - t_1(x) y + t_2(x),
\end{eqnarray}
where $t_1(x)$ is the linear trace.  If $t_1$ was not linear,
then $\deg(p) > 2$.  So $t_1(x) = a x + b$, for some $a$ and $b$
to be determined by interpolation at $x = x_0$ and $x = x_1$,
with corresponding $y$-values in $(x_0,y_{01})$, $(x_0,y_{02})$,
$(x_1,y_{11})$, and $(x_1,y_{12})$.  If for an additional sample,
at $x = x_2$ with corresponding $y$-values $(x_2,y_{21})$, $(x_2,y_{22})$,
we have $t_1(x_2) = y_{21} + y_{22}$, then we have an irreducible
quadratic factor, otherwise the two points do not lie on the same factor.

The linear trace test, called zero-sum relations, was first introduced
in~\cite{SaSuKoSa91} and further developed in~\cite{SaSaHi92,SaSa93}.
For factors of small to moderate degree, the linear trace test can be
applied in an exhaustive combinatorial enumeration as was proposed
in~\cite{GR01,GR02,Rup04}, \cite{Sas01} and improved in~\cite{Che04}.

\subsection{Parallel Algorithms}

Homotopy continuation methods are very well suited for parallel
processing as after distributing the path tracking jobs among the
computers in the network, no further communications are needed,
see~\cite{ACW89,CARW93,HW89} for granularity issues.
For computational algebraic geometry, this implies that homotopy
methods can solve much larger polynomial systems than methods in
computer algebra which are harder to adapt to
parallel computers~\cite{Ley04}.  One recent example is the
solution of the cyclic 13-roots problems with PHoM~\cite{DKK03,GKKTFM04}
for which 2,704,156 paths were tracked.

Modern homotopies in numerical algebraic geometry often appear in
a sequence like the Pieri homotopies~\cite{VW04} where the start
solutions of one homotopy lie at the end of paths defined by
another homotopy.  The homotopies to factor positive dimensional
solution sets raise job scheduling issues as the decision to
certain track paths depends on the outcome of other paths.
This paper can be regarded as a solution to the job scheduling
problems raised in~\cite{LV05}.

The parallel algorithm proposed in~\cite{LV05} exhibited a good
speedup in the path tracking jobs, but the certification with
linear traces executed in only by the master node before the
scheduling of new path tracking jobs diminished the overall
performance as all nodes were idling waiting for the assignment
of new path tracking jobs.  At the end of~\cite{LV05} we outlined
a probabilistic complexity study simulated in Maple, suggesting
various job scheduling techniques.  In this paper we report on its
parallel implementation confirming the efficiency of the approach.

\section{USING MONODROMY MORE EFFICIENTLY}

The monodromy breakup algorithm of~\cite{SVW3,SVW4}
is sketched by Algorithm~\ref{algseqmb}.  On input
is a witness set $W_L$ and on output a partition of $W_L$,
corresponding to the irreducible decomposition.

\begin{algorithm}  \label{algseqmb}
{\rm Monodromy Breakup certified by Linear Trace:
$\cal P$ = Breakup($W_L,d,N$)

\begin{tabular}{l}
  {\bf Input:} $W_L$, $d$, $N$. \\
  {\bf Output:} $\cal P$. \\
\\
  0. initialize $\cal P$ with $d$ singletons; \\
  1. generate two slices $L'$ and $L''$ parallel to $L$; \\
  2. track $d$ paths for witness set with $L'$; \\
  3. track $d$ paths for witness set with $L''$; \\
  4. {\bf for} $k$ {\bf from} 1 {\bf to} $N$ {\bf do} \\
\hspace{4mm} 4.1 generate new slices $K$ and a random~$\gamma$; \\
\hspace{4mm} 4.2 track $d$ paths to a new slice; \\
\hspace{4mm} 4.3 generate a new random~$\gamma$; \\
\hspace{4mm} 4.4 track $d$ paths to return to the base slice; \\
\hspace{4mm} 4.5 compute the permutation and update $\cal P$; \\
\hspace{4mm} 4.6 {\bf if} linear trace test certifies $\cal P$ \\
\hspace{10.5mm} {\bf then} leave the loop; \\
\hspace{9.7mm} {\bf end if}; \\
\hspace{4mm}  {\bf end for}.
\end{tabular}
}
\end{algorithm}

Our first parallel implementation of this algorithm,
described in~\cite{LV05}, uses a master/servant model
where the master node distributes the paths among the
available processors in the network.
According to our experimental results, a sizeable speedup is achieved
by distributing the routine path-tracking jobs to different nodes.
However, a probabilistic study in~\cite{LV05} suggested that we can
save some work by taking a smaller one-path-one-point tracking job as
an atomic task.

Following the previous discussion, we take two generic slices
$\LL_1$ and $\LL_2$ (in this case these are hyperplanes) and look
at the witness points $W_1$ and $W_2$. Consider the bipartite
graph with vertices $W_1$ on one side and $W_2$ on the other.

One atomic step of the monodromy breakup algorithm
consists of creating a map like $\phi_{\gamma, \LL_1, \LL_2}$.
We can visualize such a map by connecting the points of $W_1$ and $W_2$
that map into each other with an edge.

\begin{figure}[h] \label{fig-bipartite}
\input{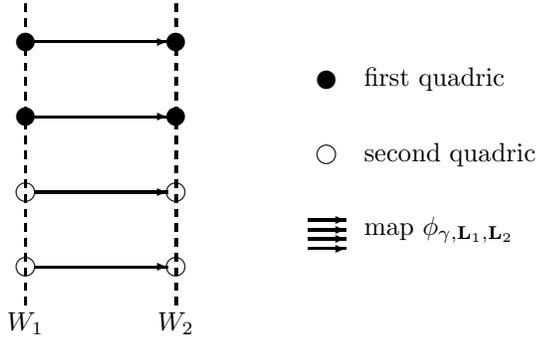}
\caption{The bipartite graph $W_1\leftrightarrow W_2$ for {\tt
cyclic4}}
\end{figure}

As you may see in our example, in order to create one permutation
we need to construct 8 edges in the graph. If one is lucky then it
may take [MBA-P] only one permutation to decompose {\tt cyclic4}:

\begin{figure}[h]\label{fig-12-34}
\input{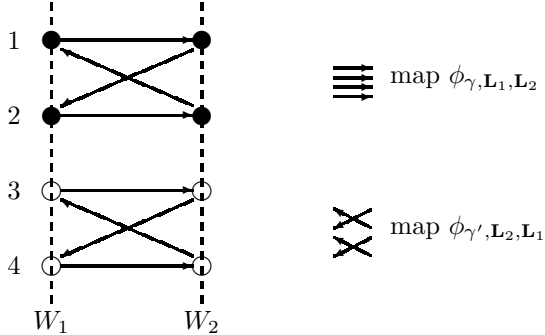}
\caption{Permutation (12)(34) for {\tt cyclic4}}
\end{figure}

The connected components of the graph in Figure \ref{fig-12-34}
correspond to the two witness sets that, in turn, represent two
irreducible components of the solution set of {\tt cyclic4}: two
quadric curves.

\begin{figure}[h]\label{fig-12-34-6edges}
\input{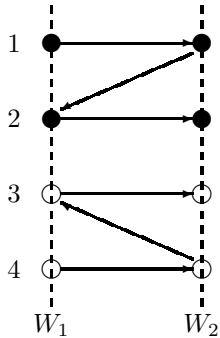}
\caption{ A 6-edge graph for {\tt cyclic4}}
\end{figure}

In fact 2 out of 8 edges in Figure \ref{fig-12-34} can be removed
keeping the connected components still connected, see Figure
\ref{fig-12-34-6edges}. Since each edge can be created by tracking
only one point of a witness set, we may avoid doing extra work by
trying to create as few edges as possible.

\section{A NEW ALGORITHM}

In this section we first describe the sequential version of
our new monodromy breakup algorithm before addressing its
parallel execution.

\subsection{Serial Version}

\begin{figure}[h]
\begin{center}
\input{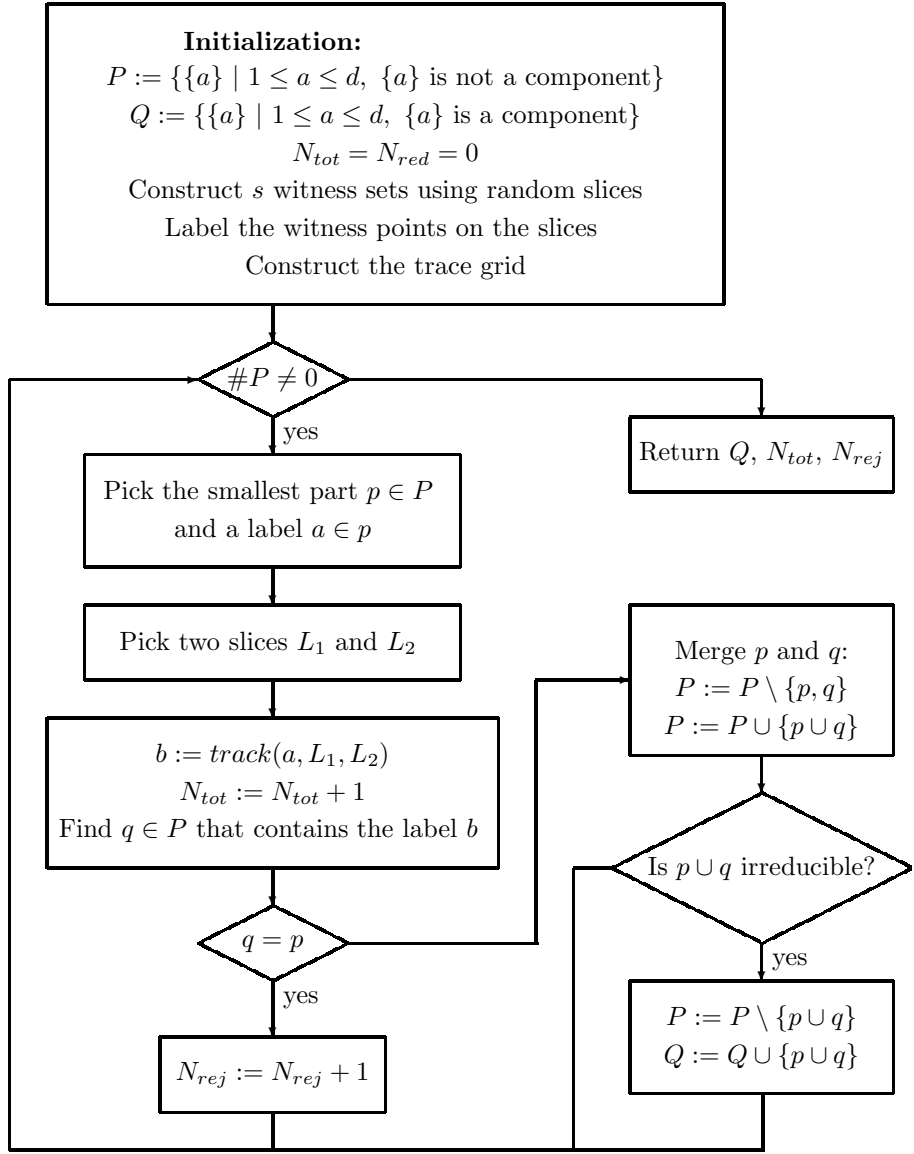}
\caption{The flow chart of our new Monodromy Breakup Algorithm}
\label{fig_flow}
\end{center}
\end{figure}

The flow chart of our new algorithm is show in Figure~\ref{fig_flow}.
As in Algorithm~\ref{algseqmb}, we also have the initialization of the
``trace grid'', which are the two witness sets on two parallel slices
needed to certify the irreducible decomposition using linear traces.
While Algorithm~\ref{algseqmb} first completes all the loops for all
points in a witness set before proceeding to the next level, our new
approach initializes $s$ new witness sets which are available for
generating loops.

The main loop of the new algorithm shown in Figure~\ref{fig_flow}
leaves much freedom to complete loops between any two slices.
For every slice, the algorithm keeps track of the number
of loops that did not yielded a permutation, stored in $N_{rej}$.
Based on these statistics, the algorithm can discriminate against
slices which were not productive in the past and select those
slices which led to more new permutations.

As explained in the previous section, the algorithm typically returns
with a certified decomposition before all loops are completed.
This is the main reason why on single processors, our new algorithm
outperforms Algorithm~\ref{algseqmb}.  At the same time, the new
algorithm is more suitable for parallel execution, as we will explain
next.

\subsection{Parallel Version}

The parallel version of our new algorithm runs in a master/servant model.
The initialization phase is very similar to the initialization of the
parallel version of Algorithm~\ref{algseqmb}, with the master distributing
path tracking jobs evenly among all nodes.  After the initialization, the
master node keeps looking for available path tracking nodes to assign
paths and the other nodes are either busy tracking paths or ready to
start new path tracking jobs.

Compared to our previous parallel algorithm described in~\cite{LV05},
the computation of the linear trace by the master node is now interleaved
by path tracking jobs performed on the other nodes.

\section{EXPERIMENTAL RESULTS}

Our algorithms are implemented using the path tracking routines
in PHCpack~\cite{Ver99}, extended in~\cite{SVW7} with facilities for
a numerical irreducible decomposition.
As in~\cite{VW04}, we apply MPI for message passing.  Our main
program is written in C and links with the interface of PHCpack.

Our equipment consists of two personal cluster machines purchased from
Rocketcalc ({\tt www.rocketcalc.com}) for a total of 12 2.4Ghz CPUs,
served by a Dell workstation with two dual 2.4Ghz processors.
So in total we have 14 processors at our disposal.

\subsection{Plain Parallel Path Tracking}

In Table~\ref{tabfirst} we show with three runs the main defect
of our first parallel implementation presented in~\cite{LV05}.
While we have no real control over the number of loops it takes
to complete the factorization, we observe from the data in
Table~\ref{tabfirst} that as the total number of loops increases,
the work done by the master node increases.

\begin{table}[hbt]
\begin{center}
\begin{tabular}{|c||c|c|c|} \hline
   \#loops & 4 & 6 & 9 \\ \hline
 min track & ~8.0~sec & 10.9~sec & 18.5~sec \\
 max track & 10.8~sec & 15.7~sec & 21.8~sec \\
 master    & ~1.8~sec & ~3.8~sec & ~7.6~sec \\ \hline
 total     & 12.6~sec & 19.5~sec & 29.4~sec \\ \hline
\end{tabular}
\caption{Three runs with the first parallel monodromy breakup algorithm,
executing respectively 4, 6, and 9 loops to factor a curve of degree~144
in 8~space on 14 processors.  We report the minimal and maximal time the
nodes spent tracking paths, and the time spent by the master node
certifying the decomposition and scheduling the jobs.}
\label{tabfirst}
\end{center}
\end{table}

Although the cyclic 8-roots system is a problem of modest size,
we already observe in Table~\ref{tabfirst} that for 9 loops,
more than 25\% of the time is spent by the master node, while all
the other nodes are idling.  For larger problems and more processors,
the poor performance of this first implementation will become even
more apparent.

\subsection{Performance of the new Algorithm}

Table~\ref{tabsecond} reports on five runs done on 14 processors to
decompose the curve of degree 144 defined by the cyclic 8-roots system.
For three of the five runs we used three new slices.
In the last two runs we see that the total time decreases if we use only
two new slices.

\begin{table}[hbt]
\begin{center}
\begin{tabular}{|c||c|c|c||c|c|} \hline
   & \multicolumn{3}{c||}{3 new slices}
   & \multicolumn{2}{c|}{2 new slices} \\ \hline
    \#runs & 1 & 2 & 3 & 4 & 5 \\ \hline
   initial & 8.73 & 9.01 & 8.89 & 6.54 & 6.98 \\ \hline
    master & 6.06 & 6.22 & 6.18 & 6.67 & 7.10 \\
 min track & 5.96 & 6.16 & 6.07 & 6.60 & 7.02 \\
 max track & 6.06 & 6.24 & 6.23 & 6.11 & 7.15 \\ \hline
     total & 14.9 & 15.4 & 15.3 & 13.4 & 14.2 \\ \hline
\end{tabular}
\caption{Five runs with our new parallel monodromy breakup algorithm,
three times with 3 new slices and two times with 2 new slices.
We report the time used for initialization, the time spent by the
master node, the minimal and maximal time for the nodes spent tracking
paths, and the total time.  All reported times are expressed in seconds.}
\label{tabsecond}
\end{center}
\end{table}

In Table~\ref{tabsecond} we see an even distribution of the time
spent by the nodes.  Using fewer slices reduces the initialization
time at the expense of a slightly higher running time in the main loop.

Comparing to the timings in Table~\ref{tabfirst} we do not notice
such a wide fluctuation in the total execution time between different
numbers of loops.  The total execution time of the most favorable
situation reported in Table~\ref{tabfirst} is only slightly lower
than the best total time in Table~\ref{tabsecond}.

Finally, we report on a calculation of a larger example, the
ideal of adjacent 2-by-2 minors of a general 2-by-9 matrix of
18 unknowns, see~\cite{DES98} and~\cite{HS00}.
This system in 18 variables defines a 10-dimensional surface of degree 256
which factors in 34 irreducible components.
The total execution time of our new monodromy breakup algorithm
on 14 processors is 97.1 seconds, of which 62.7 are spent on the
initialization, 33.8 seconds by the master node in the main loop
while the time path tracking on the other noeds fluctuated
between 32.9 and 35.6 seconds.

The performance of our first parallel algorithm on this system is
even more erratic.  The very best complete run of 3 monodromy loops
took 122.9 seconds on 14 processors, where the path tracking time
ranged between 75.9 and 104.4 seconds.  Even on this very best run,
our new algorithm still takes only 80\% of the time spent by the
first parallel monodromy breakup algorithm of~\cite{LV05}.

\section{CONCLUSIONS}

In this paper we report on the development and performance of a new
monodromy breakup algorithm.  Experimental results show a more
predictable and regular performance than our first parallel
implementation of~\cite{LV05}.  Thanks to this increased performance,
it is now possible to factor solution sets of larger degrees.

\bibliographystyle{plain}
\bibliography{monodromy}

\begin{thebibliography}{10}

\bibitem{ACW89}
D.C.S. Allison, A.~Chakraborty, and L.T. Watson.
\newblock Granularity issues for solving polynomial systems via globally
  convergent algorithms on a hypercube.
\newblock {\em J. of Supercomputing}, 3:5--20, 1989.

\bibitem{BCGW93}
C.~Bajaj, J.~Canny, T.~Garrity, and J.~Warren.
\newblock Factoring rational polynomials over the complex numbers.
\newblock {\em SIAM J. Comput.}, 22(2):318--331, 1993.

\bibitem{Bjo85}
G.~Bj\"{o}rck.
\newblock Functions of modulus one on {$Z_p$} whose {F}ourier transforms have
  constant modulus.
\newblock In J.~Szabados and K.~Tandori, editors, {\em Proceedings of the
  Alfred Haar Memorial Conference, Budapest}, volume~49 of {\em Colloquia
  Mathematica Societatis J\'anos Bolyai}, pages 193--197. North Holland, 1985.

\bibitem{Bjo89}
G.~Bj\"{o}rck.
\newblock Functions of modulus one on {$Z_n$} whose {F}ourier transforms have
  constant modulus, and ``cyclic n-roots''.
\newblock In J.S. Byrnes and J.F. Byrnes, editors, {\em Recent Advances in
  Fourier Analysis and its Applications}, volume 315 of {\em NATO Adv. Sci.
  Inst. Ser. C: Math. Phys. Sci.}, pages 131--140. Kluwer, 1989.

\bibitem{CARW93}
A.~Chakraborty, D.C.S. Allison, C.J. Ribbens, and L.T. Watson.
\newblock The parallel complexity of embedding algorithms for the solution of
  systems of nonlinear equations.
\newblock {\em IEEE Transactions on Parallel and Distributed Systems},
  4(4):458--465, 1993.

\bibitem{Che04}
G.~Ch{\`{e}}ze.
\newblock Absolute polynomial factorization in two variables and the knapsack
  problem.
\newblock In J.~Gutierrez, editor, {\em Proceedings of the 2004 International
  Symposium on Symbolic and Algebraic Computation (ISSAC 2004)}, pages 87--94.
  ACM, 2004.

\bibitem{CG05}
G.~Ch{\`{e}}ze and A.~Galligo.
\newblock Four lectures on polynomial absolute factorization.
\newblock In A.~Dickenstein and I.Z. Emiris, editors, {\em Solving Polynomial
  Equations: Foundations, Algorithms, and Applications}, volume~14 of {\em
  Algorithms and Computation in Mathematics}, pages 339--392. Springer--Verlag,
  2005.

\bibitem{CGKW02}
R.M. Corless, A.~Galligo, I.S. Kotsireas, and S.M. Watt.
\newblock A geometric-numeric algorithm for factoring multivariate polynomials.
\newblock In T.~Mora, editor, {\em Proceedings of the 2002 International
  Symposium on Symbolic and Algebraic Computation (ISSAC 2002)}, pages 37--45.
  ACM, 2002.

\bibitem{CGHKW01}
R.M. Corless, M.W. Giesbrecht, M.~van Hoeij, I.S. Kotsireas, and S.M. Watt.
\newblock Towards factoring bivariate approximate polynomials.
\newblock In B.~Mourrain, editor, {\em Proceedings of the 2001 International
  Symposium on Symbolic and Algebraic Computation (ISSAC 2001)}, pages 85--92.
  ACM, 2001.

\bibitem{DKK03}
Y.~Dai, S.~Kim, and M.~Kojima.
\newblock Computing all nonsingular solutions of cyclic-n polynomial using
  polyhedral homotopy continuation methods.
\newblock {\em J. Comput. Appl. Math.}, 152(1-2):83--97, 2003.

\bibitem{DvH01}
B.~Deconinck and M.~van Hoeij.
\newblock Computing {R}iemann matrices of algebraic curves.
\newblock {\em Physica D}, 152:28--46, 2001.

\bibitem{DES98}
P.~Diaconis, D.~Eisenbud, and B.~Sturmfels.
\newblock Lattice walks and primary decomposition.
\newblock In B.E. Sagan and R.P. Stanley, editors, {\em Mathematical Essays in
  Honor of Gian-Carlo Rota}, volume 161 of {\em Progress in Mathematics}, pages
  173--193. Birkh\"auser, 1998.

\bibitem{GR01}
A.~Galligo and D.~Rupprecht.
\newblock Semi-numerical determination of irreducible branches of a reduced
  space curve.
\newblock In B.~Mourrain, editor, {\em Proceedings of the 2001 International
  Symposium on Symbolic and Algebraic Computation (ISSAC 2001)}, pages
  137--142. ACM, 2001.

\bibitem{GR02}
A.~Galligo and D.~Rupprecht.
\newblock Irreducible decomposition of curves.
\newblock {\em J.\ Symbolic Computation}, 33(5):661--677, 2002.

\bibitem{GKMYZ04}
X.-S. Gao, E.~Kaltofen, J.~May, Z.~Yang, and L.~Zhi.
\newblock Approximate factorization of multivariate polynomials via
  differential equations.
\newblock In J.~Gutierrez, editor, {\em Proceedings of the 2004 International
  Symposium on Symbolic and Algebraic Computation (ISSAC 2004)}, pages
  167--174. ACM, 2004.

\bibitem{GKKTFM04}
T.~Gunji, S.~Kim, M.~Kojima, A.~Takeda, K.~Fujisawa, and T.~Mizutani.
\newblock {PHoM} -- a polyhedral homotopy continuation method for polynomial
  systems.
\newblock {\em Computing}, 73:55--77, 2004.

\bibitem{HW89}
S.~Harimoto and L.T. Watson.
\newblock The granularity of homotopy algorithms for polynomial systems of
  equations.
\newblock In G.~Rodrigue, editor, {\em Parallel processing for scientific
  computing}, pages 115--120. SIAM, 1989.

\bibitem{HS00}
S.~Ho\c{s}ten and J.~Shapiro.
\newblock Primary decomposition of lattice basis ideals.
\newblock {\em Journal of Symbolic Computation}, 29(4\&5):625--639, 2000.

\bibitem{HWSZ00}
Y.~Huang, W.~Wu, H.J. Stetter, and L.~Zhi.
\newblock Pseudofactors of multivariate polynomials.
\newblock In C.~Traverso, editor, {\em Proceedings of the 2000 International
  Symposium on Symbolic and Algebraic Computation (ISSAC 2000)}, pages
  161--168. ACM, 2000.

\bibitem{Kal00}
E.~Kaltofen.
\newblock Challenges of symbolic computation: my favorite open problems.
\newblock {\em J.\ Symbolic Computation}, 29(6):891--919, 2000.

\bibitem{Ley04}
A.~Leykin.
\newblock On parallel computation of {G}r{\"{o}}bner bases.
\newblock In Y.~Yang, editor, {\em Proceedings of the 2004 International
  Conference on Parallel Processing Workshops, 15-18 August 2004, Montreal,
  Quebec, Canada. High Performance Scientific and Engineering Computing}, pages
  160--164. IEEE Computer Society, 2004.

\bibitem{LV05}
A.~Leykin and J.~Verschelde.
\newblock Factoring solution sets of polynomial systems in parallel.
\newblock In Tor Skeie and Chu-Sing Yang, editors, {\em Proceedings of the 2005
  International Conference on Parallel Processing Workshops. 14-17 June 2005.
  Oslo, Norway. High Performance Scientific and Engineering Computing}, pages
  173--180. IEEE Computer Society, 2005.

\bibitem{Rup04}
D.~Rupprecht.
\newblock Semi-numerical absolute factorization of polynomials with integer
  coefficients.
\newblock {\em J.\ Symbolic Computation}, 37(5), 2004.

\bibitem{Sas01}
T.~Sasaki.
\newblock Approximate multivariate polynomial factorization based on zero-sum
  relations.
\newblock In B.~Mourrain, editor, {\em Proceedings of the 2001 International
  Symposium on Symbolic and Algebraic Computation (ISSAC 2001)}, pages
  284--291. ACM, 2001.

\bibitem{SaSaHi92}
T.~Sasaki, T.~Saito, and T.~T.~Hilano.
\newblock Analysis of approximate factorization algorithm i.
\newblock {\em Japan J.\ of Industrial and Applied Math.}, 9:351--368, 1992.

\bibitem{SaSa93}
T.~Sasaki and M.~Sasaki.
\newblock A unified method for multivariate polynomial factorizations.
\newblock {\em Japan J.\ of Industrial and Applied Math.}, 10:21--39, 1993.

\bibitem{SaSuKoSa91}
T.~Sasaki, M.~Suzuki, M.~Kol\'ar, and M.~Sasaki.
\newblock Approximate factorization of multivariate polynomials and absolute
  irreducibility testing.
\newblock {\em Japan J.\ of Industrial and Applied Math.}, 8:357--375, 1991.

\bibitem{SV00}
A.J. Sommese and J.~Verschelde.
\newblock Numerical homotopies to compute generic points on positive
  dimensional algebraic sets.
\newblock {\em J.\ of Complexity}, 16(3):572--602, 2000.

\bibitem{SVW1}
A.J. Sommese, J.~Verschelde, and C.W. Wampler.
\newblock Numerical decomposition of the solution sets of polynomial systems
  into irreducible components.
\newblock {\em SIAM J.\ Numer.\ Anal.}, 38(6):2022--2046, 2001.

\bibitem{SVW3}
A.J. Sommese, J.~Verschelde, and C.W. Wampler.
\newblock Using monodromy to decompose solution sets of polynomial systems into
  irreducible components.
\newblock In C.~Ciliberto, F.~Hirzebruch, R.~Miranda, and M.~Teicher, editors,
  {\em Application of Algebraic Geometry to Coding Theory, Physics and
  Computation}, pages 297--315. Kluwer Academic Publishers, 2001.
\newblock Proceedings of a NATO Conference, February 25 - March 1, 2001, Eilat,
  Israel.

\bibitem{SVW4}
A.J. Sommese, J.~Verschelde, and C.W. Wampler.
\newblock Symmetric functions applied to decomposing solution sets of
  polynomial systems.
\newblock {\em SIAM J.\ Numer.\ Anal.}, 40(6):2026--2046, 2002.

\bibitem{SVW7}
A.J. Sommese, J.~Verschelde, and C.W. Wampler.
\newblock Numerical irreducible decomposition using {PHC}pack.
\newblock In M.~Joswig and N.~Takayama, editors, {\em Algebra, Geometry, and
  Software Systems}, pages 109--130. Springer--Verlag, 2003.

\bibitem{SVW8}
A.J. Sommese, J.~Verschelde, and C.W. Wampler.
\newblock Numerical factorization of multivariate complex polynomials.
\newblock {\em Theoretical Computer Science}, 315(2-3):651--669, 2004.
\newblock Special Issue on Algebraic and Numerical Algorithms edited by I.Z.
  Emiris, B. Mourrain, and V.Y. Pan.

\bibitem{SW96}
A.J. Sommese and C.W. Wampler.
\newblock Numerical algebraic geometry.
\newblock In J.~Renegar, M.~Shub, and S.~Smale, editors, {\em The Mathematics
  of Numerical Analysis}, volume~32 of {\em Lectures in Applied Mathematics},
  pages 749--763. AMS, 1996.
\newblock Proceedings of the AMS-SIAM Summer Seminar in Applied Mathematics.
  Park City, Utah, July 17-August 11, 1995, Park City, Utah.

\bibitem{SW05}
A.J. Sommese and C.W. Wampler.
\newblock {\em The Numerical solution of systems of polynomials arising in
  engineering and science}.
\newblock World Scientific, 2005.

\bibitem{Ver99}
J.~Verschelde.
\newblock Algorithm 795: {PHC}pack: A general-purpose solver for polynomial
  systems by homotopy continuation.
\newblock {\em ACM Trans. Math. Softw.}, 25(2):251--276, 1999.
\newblock Software available at {\tt http://www.math.uic.edu/{\~{}}jan}.

\bibitem{VW04}
J.~Verschelde and Y.~Wang.
\newblock Computing feedback laws for linear systems with a parallel {P}ieri
  homotopy.
\newblock In Y.~Yang, editor, {\em Proceedings of the 2004 International
  Conference on Parallel Processing Workshops, 15-18 August 2004, Montreal,
  Quebec, Canada. High Performance Scientific and Engineering Computing}, pages
  222--229. IEEE Computer Society, 2004.

\end{thebibliography}

\end{document}